\documentclass[useAMS,usenatbib]{mn2e}
\usepackage{times}
\usepackage{txfonts}
\usepackage{epsfig}

\newcommand{\eg}{e.g. }
\newcommand{\cgs}{erg s$^{-1}$ cm$^{-2}~$}

\newcommand{\thirt}{$13^H~$}
\newcommand{\xmm}{XMM-{\it Newton} }
\newcommand{\xmmcomma}{XMM-{\it Newton}, }
\newcommand{\chandra}{{\it Chandra} }
\newcommand{\rosat}{{\it ROSAT} }
\newcommand{\einstein}{{\it Einstein} }

\newcommand{\spitzer}{{\it Spitzer }}

\title[Constraints on the distribution of absorption in the X-ray selected AGN population]{Constraints on the distribution of absorption in the X-ray selected AGN population found in the \thirt XMM-{\it Newton}/{\it Chandra} deep field}
\author[Dwelly]{T. Dwelly$^{1}$\thanks{E-mail:td@mssl.ucl.ac.uk}, M.J. Page$^{1}$, N.S. Loaring$^{1}$, K.O. Mason$^{1}$, I. McHardy$^{2}$, K. Gunn$^{2}$, \and T. Sasseen$^{3}$ \\
$^{1}$Mullard Space Science Laboratory, UCL, Holmbury St. Mary, Dorking, Surrey, RH5 6NT, UK.\\
$^{2}$Dept. of Physics, University of Southampton, Southampton, SO17 1BJ, UK.\\
$^{3}$Department of Physics, UC Santa Barbara, Santa Barbara, Ca., 93106, USA.}

\begin{document}

\date{\today}

\pagerange{\pageref{firstpage}--\pageref{lastpage}} \pubyear{2005}

\maketitle

\label{firstpage}

\begin{abstract}
We present an analysis of the X-ray properties of sources detected in the \thirt \xmm deep (200ks) field.
In order to constrain the absorbed AGN population, we use extensive Monte Carlo simulations to directly compare the X-ray colours of observed sources with those predicted by several model distributions.
In particular, we have carried out our comparisons over the entire 0.2--10 keV energy range of the \xmm cameras, making our analysis sensitive to a large range of absorbing column densities.
We have tested the simplest form of the unified scheme, whereby the intrinsic luminosity function of absorbed AGN is set to be the same as that of their unabsorbed brethren, coupled with various model distributions of absorption.
Of the tested distributions, the best fitting model has the fraction of AGN with absorbing column $N_H$, proportional to (log$N_H$)$^8$.
We have also tested two extensions to the unified scheme: an evolving absorption scenario, in which the fraction of absorbed sources is larger at higher redshifts, and a luminosity dependent model in which high luminosity AGN are less likely to be absorbed.
Both of these models provide poorer matches to the observed X-ray colour distributions than the best fitting simple unified model.
We find that a luminosity dependent density evolution luminosity function reproduces poorly the 0.5--2 keV source counts seen in the \thirt field.
Field to field variations could be the cause of this disparity.
Computing the X-ray colours with a simple absorbed power-law spectral model is found to over-predict, by a factor of two, the fraction of hard sources that are completely absorbed below 0.5 keV, implying that an additional source of soft-band flux must be present in a number of the absorbed sources.
The tested synthesis models predict that around 16\% of the detections in the \thirt field are due to AGN at $z>3$. 
However, so far, only a single AGN with $z > 3$ has been identified in our approximately $50\%$ complete optical spectroscopy follow up program.
Finally, we use our simulations to demonstrate the efficacy of a hardness ratio selection scheme at selecting absorbed sources for further study.
Using this selection scheme, we show that around 40\% of the \thirt sample are expected to be AGN with $N_H > 10^{22}$ cm$^{-2}$.

\end{abstract}


\section{Introduction}

Deep X-ray surveys have progressively resolved an increasing fraction of the soft X-ray background (XRB), into faint point-like sources.
Most recently, the ultra-deep \chandra surveys (\citealt{rosati02}, \citealt{brandt01}), have resolved over $90\%$ of the 0.5--2 keV XRB.
Source counts have been measured in these fields to limiting fluxes of a few times $10^{-17} $\cgs, with corresponding sky densities of over $10^4$ sources deg$^{-2}$, \citep{alexander03}. 
Optical identification of faint X-ray sources reveals a heterogeneous mixture of objects, with the dominant class being active galactic nuclei (AGN), (\citealt{page03}, \citealt{barger03}).
These X-ray selected AGN have a range of luminosities spanning several orders of magnitude, are found at redshifts up to 5, and exhibit a wide range of observational properties.
The unification scheme of \citet{antonucci93} attributes the wide variety of radio, optical and X-ray characteristics seen between AGN classes, to the geometry and relative orientation of a dusty torus surrounding the central black hole.
The dusty torus model is used to to explain both the absence of broadened optical lines, and the strong X-ray absorption seen in many AGN.
The simplest version of the unified scheme is one in which the geometry of the inner regions, and hence the distribution of absorbing column densities ($N_H$), is independent of all other AGN properties.  
A number of refinements to this model have been suggested, in order to explain recent observations which are at odds with the simple unified scheme. 

The demographics of the 0.5--2 keV AGN population have been well measured, albeit to relatively bright flux limits, using \rosat/\einstein samples, for example by \citet{maccacaro91}, \citet{boyle94}, \citet{page97} and \citet{miyaji00}.
Each of these studies indicates strong evolution of the AGN X-ray luminosity function (XLF), with a peak in AGN activity at $1 < z < 2$, although the best description of this evolution is disputed.
The soft XLF is characterised by a double power-law with a knee at $L_{0.5-2} \sim 10^{44}$ erg s$^{-1}$. 
The bulk of the XRB is the result of the integrated emission of AGN having luminosities in the vicinity of this knee.

The spectral slope of the 0.1--1 keV XRB is $\Gamma \sim 2.0$ \citep{miyaji98}, approximately matching that of a typical soft band detected AGN, $\overline{\Gamma} = 1.9$, (\citealt{page03}, \citealt{piconcelli03}, \citealt{mateos05}). 
However, at harder X-ray energies the XRB flattens dramatically, and has $\Gamma \sim 1.4$ in the 1--10 keV band (e.g. \citealt{miyaji98}, \citealt{lumb02}, \citealt{deluca04}), and cannot be produced by a simple superposition of canonical $\Gamma = 1.9$ AGN spectra. 
Clearly, some additional sources of hard X-rays must exist, and a large population of heavily absorbed AGN is postulated to fill this role. 
Population synthesis models have been formulated, such as those of \citet{madau94}, \citet{comastri95}, and \citet{gilli01}, and have been successful in reproducing both the broad band spectrum of the XRB, as well as the AGN source counts observed below 10 keV.
\citet{gilli01} predict that the majority ($\geq 80\%$), of the AGN population is heavily absorbed.

A number of deep surveys of the 2--10 keV sky have been performed with \chandra and \xmm (e.g. \citealt{brandt01}, \citealt{giacconi02}, \citealt{hasinger01}).
In the 2Ms observations of the \chandra Deep Field-North (CDF-North), sources have been detected to a limiting flux of $\sim 1.4 \times 10^{-16}$ \cgs in the 2--10 keV band \citep{alexander03}.
The 1Ms observations of the \chandra Deep Field-South (CDF-South), are estimated to have resolved more than $85\%$ of the 2--10 keV XRB \citep{rosati02}.
However, due to the relatively large uncertainty ($\sim 20\%$) in the total intensity of the 2--10 keV cosmic XRB, the precise resolved fraction is still somewhat unknown \citep{deluca04}. 
It has also been shown that extrapolating the source counts seen in the Chandra deep fields to much lower fluxes does not reproduce fully the total level of the cosmic XRB, suggesting the existence of an additional very faint X-ray population (e.g. \citealt{moretti03}, \citealt{deluca04}).

The luminosity function of AGN that are selected in the 2--10 keV band, has been measured by several studies.
\citet{cowie03} demonstrated that strong evolution of the hard XLF has occurred between the $z=2-4$ and $z=0.1-1$ epochs. 
\citet{ueda03} used a sample of 247 AGN, including some from the CDF-North, to show that the XLF is best represented by a complex luminosity dependent density evolution (LDDE) model.
It should be noted that this sample does not reach to the limiting flux of the CDF-N data, but to $S_{2-10} \sim 4 \times 10^{-15}$ \cgs, where the optical identifications are reasonably complete.

The major stumbling block in understanding the nature of the faint, hard X-ray selected AGN population, is the difficulty of obtaining optical spectroscopic identifications.
The soft X-ray selected samples used for XLF determinations, for example \citealt{page97} and \citealt{miyaji00}, are primarily, or wholly, composed of bright ($R<22$), broad-line AGN counterparts, which are relatively easy to optically identify.
At fainter X-ray fluxes, such as those probed in the CDF-North, AGN without broad lines, together with normal galaxies, make up a large fraction of the identified objects \citep{barger03}.
Those non-broad-line AGN having spectroscopic identifications are almost all found with $z < 1$, in contrast to the peak of the broad-line sample, which lies at $1<z<2$. 
The large numbers of $z>1$ type-II quasars (having log$L_X > 44$, log$N_H > 22$), predicted by synthesis models have not been detected in these surveys. 
The obvious conclusion to be drawn is that the absorbed and unabsorbed AGN are taken from separate populations, a direct contradiction of the simplest unified scheme.
However, the optical follow up programs in these deep \chandra fields are by no means complete to the faint X-ray limit.
For example, in the CDF-North, only 55\% of the X-ray detections have optical counterparts with $R \leq 24$ \citep{barger03}.
Objects fainter than this limit are practically unidentifiable with current optical spectroscopic techniques.
There is a wide range of X-ray to optical flux ratios, $f_X / f_{opt}$ in these samples, and so the unidentifiable objects are not necessarily the faintest X-ray sources, and produce a significant fraction of the XRB.
The nature of these optically faint, hard X-ray objects is still not well understood.
They could be narrow line AGN at $z > 1$, with their strongest emission lines shifted out of the optical band.
A number of optically-faint \chandra sources have been identified from their near IR properties to be AGN located in luminous, evolved host galaxies at high redshifts (e.g. \citealt{cowie01}).
Alternatively, the unidentifiable sources may be AGN embedded in optically thick dusty galaxies at moderate redshifts, the faintness of the hosts precluding identification \citep{fabian98,severgnini03}. 

Optical studies of nearby Seyfert galaxies have found that the ratio, R, of type 2 to type 1 Seyferts, is approximately 4 \citep{maiolino95}.
The hard X-ray study of these type 2 Seyferts by \citet{risaliti99} discovered a wide distribution of absorbing columns, but with $\sim 75\%$ of the AGN having $N_H > 10^{23}$ cm$^{-2}$.  
However, this study was limited to the very local universe, $\langle d \rangle = 24$ Mpc, and to low nuclear luminosities, $M_B > -22 $; the behaviour in the rest of the redshift-luminosity plane is less well understood. 
The distribution of absorption in X-ray selected AGN is poorly constrained, the prime difficulty being that the greater an object's $N_H$, the lower is its chance of being detected, or optically identified.
There have been several published cases of AGN in which the absorbing columns inferred by optical and X-ray measurements differ significantly (e.g. \citealt{maiolino01}, \citealt{page01}, \citealt{loaring03}).
Despite these limitations, the $N_H$ distribution, for hard X-ray selected AGN, has been estimated, for relatively bright samples, by \citet{ueda03}.
These authors have primarily determined $N_H$ in their optically identified sample by examination of X-ray hardness ratio between the 0.5--2 and 2--10 keV bands.
The distribution of absorption within the sample is described with a luminosity dependent $N_H$ model, in which high intrinsic luminosity AGN are less likely to be heavily absorbed.
This model does require some additional Compton thick AGN to reproduce fully the XRB when extrapolated to harder energies.

So, despite the progress made in resolving, and to some extent optically identifying, the hard X-ray population, it has still not been possible to delimit the distribution of absorption in AGN.
This problem is particularly acute for the heavily absorbed, high-$z$ AGN; few of which have been detected and identified.
However, by better constraining the $f(N_H)$ in faint AGN, we can hope to answer many questions about the geometry, composition and evolution of the dusty torus.
For example, the strength of the luminosity dependence of $f(N_H)$ can tell us about how the radiation from the accretion disk influences the surrounding torus, and/or how the torus geometry scales with black hole mass.
If some redshift evolution of $f(N_H)$ is detected, is it related to the overall evolution of the AGN luminosity function? 
Does the dusty torus form coevally with the black hole, and is the amount of absorbing material related to the black hole mass?

In this study we use X-ray hardness ratios as an indicator of absorption in the spectra of the sources in our sample.
Many authors (e.g. \citealt{mainieri02}, \citealt{dellaceca04}, and \citealt{perola04}) have shown that colour based analyses are effective in deriving the properties of \xmm sources which are detected with too few counts to permit full spectral fitting.
In these optically identified samples, the AGN with and without broad emission lines are seen to occupy separate regions in X-ray colour-colour plots.

We present in this paper an analysis of the X-ray properties of sources detected by \xmm in the \thirt deep field.
In section \ref{section_observations} we describe the \xmm observations.
In order to escape the possible biases introduced by the incompleteness of optical identification programmes, we have devised a method to probe the $f(N_H)$ of our sample.
Our technique does not depend on optical identification of the sample, permitting the inclusion of the optically faint X-ray detections. 
The simulations use a model XLF to describe the intrinsic distribution of all AGN in redshift and (de-absorbed) luminosity space; this is coupled to a model $N_H$ function, to generate a synthetic AGN population (described in section \ref{section_simulation_method}).
We simulate how this model population would be seen with \xmmcomma accounting for both the selection function caused by the complex EPIC detector imaging characteristics, and the nuances of the source detection process (see section \ref{section_capabilities}).
The output products of the simulation allow direct comparison of each of the model $N_H$ distributions with the \thirt sample.
We then compare the predictions of several simple unified scheme models of the $N_H$ distribution, by using a statistical comparison of the X-ray colour distributions found in the data and models (section \ref{section_colour_stats}). 
Furthermore, we test two examples of more complex $N_H$ distribution models taken from the literature, and compare them to the \thirt sample. 
In section \ref{section_source_counts} we compare the source counts found in the \thirt field and those predicted by the model simulations.
Finally, in section \ref{section_discussion} we discuss our results and their implications for AGN torus models, and for the evolving XLF model of \citet{miyaji00}.

Throughout the paper we use a lambda-dominated flat cosmology with $H_0 = 70$ km s$^{-1}$Mpc$^{-1}$, $(\Omega_M,\Omega_{\Lambda}) = (0.3,0.7)$. 
$L_{Emin-Emax}$ refers to an object's de-absorbed X-ray luminosity in the observed $E_{min}-E_{max}$ band.
$N_H$ is the equivalent hydrogen column density in units of cm$^{-2}$.
We refer to the $N_H$ distribution function as $f(N_H)$, and define it to be the fraction of all AGN, per unit log$N_H$, which have absorbing column $N_H$.
We define a power-law spectrum to be $F \propto E^{-\Gamma}$, where $F$ is the flux in units of photons keV$^{-1}$ s$^{-1}$ cm$^{-2}$, $E$ is the photon energy in keV, and $\Gamma$ is the photon index. 

\section[]{Models of the distribution of absorption in AGN}
\label{section_absorption_models}
The unified model attributes the X-ray absorption seen in AGN to a dusty torus surrounding the central super massive black hole \citep{antonucci93}.
For certain orientations, the torus obscures the observer's line of sight to the X-ray/UV emitting accretion disk. 
There have been a wide range of absorbing columns inferred from the X-ray spectra of various AGN, ranging from effectively zero absorption, to column densities over $10^{25}$cm$^{-2}$.
The unified scheme states that all AGN are intrinsically similar, and that the observational differences between the various AGN types are due to the orientation of the observer.
Therefore, it is the geometry of the dust torus which determines the amount of obscuring material along the observer's line of sight to the central X-ray emitting regions.
If we assume that all AGN have the same geometry, then it is only the properties of the torus which determine the observed $f(N_H)$ in the AGN population as a whole.
A typical zeroth order approach is to speculate that this characteristic geometry is independent of the luminosity of the central engine, and has not evolved over cosmological timescales.  
However, alternative scenarios are postulated, for example by \citet{gilli01} and \citet{ueda03}, which imply more complex forms for $f(N_H)$.
The \citet{gilli01} model suggests that some evolution of the average torus properties has occurred over cosmological timescales, with more absorbed AGN at high redshifts.
The \citet{ueda03} model predicts that the geometry, specifically the opening angle, of the obscuring torus is determined by the luminosity of the central engine.

In this study, we compare the predictions of several different forms of $f(N_H)$. 
A very simple description of $f(N_H)$, is a continuous distribution, in which the number of AGN per unit log$N_H$ is proportional to $($log$N_H )^{\beta}$, over the range $19~<$~log~$N_H~<~25$.
A similar parameterisation was adopted in the synthesis model of \citet{gandhi03}, who found that setting $\beta = 2,5$ or $8$ gave acceptable fits to the XRB (it should be noted that the authors used a separately evolving luminosity function for absorbed AGN).
We have tested three such $f(N_H)$, and refer to them as the $\beta =2$, $\beta =5$ and $\beta =8$ models.

Model A of \citet{gilli01} combines the $f(N_H)$ observed in the optically selected Seyfert-2 galaxies \citep{risaliti99}, with a fixed ratio, $R$, of absorbed to unabsorbed AGN. 
We have tested two similar $f(N_H)$ models here, where $R$ is constant and set to 4, and 8, and refer to these as the $R=4$, and $R=8$ models respectively.
The measured distribution of \citet{risaliti99} contains a number of AGN where only the lower or upper limit on absorption is known.
So, for the purposes of our study, all those absorbed AGN having log$N_H < 22$ are evenly distributed in the $21 <$ log$N_H < 22$ interval, and those with log$N_H > 25$ are set to have log$N_H = 25$.
We also compare the predictions of Model B of \citealt{gilli01}, which is similar to Model A, above, but with $R=4$ at $z=0$ increasing to $R=10$ for $z \ge 1.32$; we refer to this as the $R=R(z)$ model. 

\begin{figure}
\includegraphics[angle=270,width=80mm]{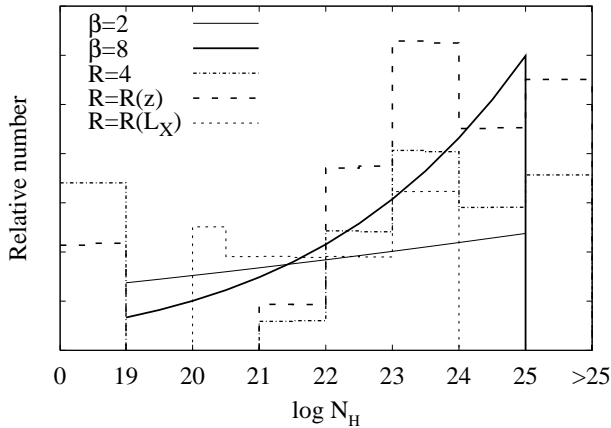}
\caption{The distribution of absorbing column densities for a subset of the tested models. AGN with log$N_H < 19$ are shown in the leftmost bin, and those with log $N_H > 25$ in the rightmost bin.}
\label{nh_histo}
\end{figure}

In addition, we test the luminosity dependent $f(N_H)$ function of \citealt{ueda03}, in which high luminosity AGN are more likely to have lower absorbing columns.
We have converted from our observed frame $L_{0.5-2}$ to rest-frame $L_{2-10}$ using the specific spectral slope, and redshift of each simulated AGN; we refer to this as the $R = R(L_X)$ model.
The $R = R(L_X)$ model distribution does not include any AGN having absorbing columns outside $20 <$ log$N_H < 24$.
Finally, we employ a zero absorption scenario to provide a base-line to the more realistic models; we call this the $R=0$ model.
A subset of the  model $f(N_H)$ are shown in fig. \ref{nh_histo}.

To reiterate, for all the tested models, we take only the $f(N_H)$ part from the published model, and always use the LDDE1 model XLF of \citet{miyaji00} to describe both absorbed and unabsorbed AGN.

Recent X-ray spectral fitting analyses have found that after absorption is considered, the mean AGN photon index $\Gamma$ is $\sim 1.9$ (\citealt{piconcelli03}, \citealt{page03}, \citealt{mateos05}). 
The absorbed and unabsorbed AGN show similar mean slopes, and there is no significant evolution seen even up to $z = 5$.
However, there is still an intrinsic scatter of slopes about the mean, and this will have some effect on the observed colours and/or detectability of sources.
Therefore, we have used a Gaussian distribution of slopes $g(\Gamma)$, to represent the spectra of the simulated AGN, with $\overline{\Gamma} = 1.9$, and $\sigma_{\Gamma} = 0.2$.
We have not considered sources with slopes outside the range $1.2 < \Gamma < 2.6$. 
\citet{piconcelli03} found no apparent dependence of de-absorbed $\Gamma$ on $z$, $N_H$, or flux, therefore we assume that $g(\Gamma)$ is independent of the other AGN spectral parameters.

\section[]{Observations}
\label{section_observations}
The \xmm data consist of three observations of the \thirt field totalling 200 ks, of which $\sim 120$ks is unaffected by soft proton flaring.
This field was the location of one of the deepest \rosat surveys \citep{mchardy98}, due to its unusually low Galactic absorbing column ($N_H \sim 8 \times 10^{19}$cm$^{-2}$).
In addition, the \thirt field has been the subject of a host of multi-wavelength observations, including a mosaic of four 30ks \chandra pointings covering the \xmm field of view \citep{mchardy03}, and extensive, very deep radio mapping with the VLA \citep{seymour04}. 

The data from the European Photon Imaging Cameras (EPIC) were reduced with version 6.0 of the Science Analysis Software (SAS) task-set to produce images in four energy bands (0.2--0.5, 0.5--2, 2--5, and 5--10 keV).
Our source detection process uses the standard SAS tasks \textsc{EBOXDETECT} and \textsc{EMLDETECT} together with a custom background fitting task. 
We perform the source detection routines on the combined data from all three (MOS1, MOS2 and pn) EPIC detectors.
We used simulations to determine detection likelihood limits such that we expect only 3\% of the final sourcelist to be spurious detections.
Using these likelihood limits we detected 225 sources. 
The approximate limiting fluxes (in units of $10^{-15}$ \cgs), are 0.5 (0.2--0.5 keV), 0.5 (0.5--2 keV), 1.2 (2--5 keV), and 5 (5--10 keV).
In the 2--5 keV band our sample reaches a factor $\sim 10$ fainter than the knee of the source counts, where the contribution to the XRB, per unit log flux, is greatest.
A full description of the data reduction, source searching, and detection threshold determination processes is given in \citet{loaring05}.

The purpose of this study is to test the predictions of a number of model AGN populations against AGN in the \thirt field.
However, we do expect to find a small number of non-AGN sources in the full X-ray sample, and these could affect our comparisons.
Our ongoing optical spectroscopic follow-up program has identified counterparts to over 100 of the \xmm sources. 
In particular, the brightest $(R<22)$ optical counterparts are $92\% (81/88)$ identified.
Four of the sources, including the brightest source in the field, are associated with foreground stars, and therefore are not included in this study.
We do not expect many of the remaining optically faint, $R>22$, counterparts to be identified with stars.
\textsc{EMLDETECT} finds four X-ray sources with high likelihood of being extended, and with FWHM~$>~16$\arcsec.
Of these four sources, three were identified as clusters by \citet{jones02} in an analysis of the \rosat imaging of the \thirt field.
\citet{jones02} found two additional clusters in the \thirt field, however, these are both located at the very edge of the EPIC field of view, (where the vignetting is most pronounced), and are not detected as extended sources by \textsc{EMLDETECT}. 
Our AGN population models do not include clusters of galaxies, so we discount the four extended sources from our analysis.
A small number of fainter clusters are expected to remain in the sample, but not be flagged as extended by \textsc{EMLDETECT}.
We estimate this number by extrapolating the $N(>S_{0.5-2})$ plot of \citet{jones02} to lower fluxes, whilst incorporating the effective area determination of our survey \citep{loaring05}.
Assuming that the flux limit for detecting extended sources is twice that for point sources, and using a $N(>S_{0.5-2})$ slope of $0.5$, we predict that approximately five additional clusters will remain in the sample.

After the stars and obvious clusters are removed, the resulting \xmm sample contains a total of 217 sources of which the vast majority are likely to be AGN.

\section[]{Simulation method}
\label{section_simulation_method}
We have devised a Monte Carlo simulation technique which allows direct comparison of the pattern of X-ray colours produced by AGN absorption models, with the pattern seen in the \thirt sample. 
For this, we have extended the \xmm imaging simulation method of \citet{loaring05}, to a multi-band approach.
We model the EPIC point spread function, vignetting and diffuse background in the same way as before.
This method accounts for observational biases and the complex selection function at work in the sample.
Each iteration of the simulation method consists of four steps.
i) We generate an input source population, with each member having a set of randomly distributed parameters $L_X,z,N_H$ and $\Gamma$, from the XLF, $f(N_H)$, and $g(\Gamma)$ models. 
ii) Multi-band count rates are calculated for each simulated source according to, $L_X,z,N_H,\Gamma$, and the chosen spectral model.
iii) Random source positions are assigned, and the sourcelist is folded through a model of the EPIC imaging response to create multi-band images for each EPIC detector.
iv) A source-detection chain is carried out on the resulting multi-band images to create an output sourcelist.
We repeat these steps for 100 simulated fields for each $f(N_H)$ model, and for two AGN spectral models.
Simulated images are produced separately for the MOS1, MOS2 and pn cameras, then combined to produce a single image in each of the four energy bands for source searching purposes.

\subsection[]{Modelling the AGN population}
To generate the catalogue for the \thirt field, \citet{loaring05} used model $N(>S)$ curves to represent the AGN population independently in each of four energy bands.
While valid for monochromatic studies, this technique is not suitable for colour analyses, since it takes no account of the multi-band properties of individual sources.
We assume that there exists a single {\em intrinsic} XLF which describes all AGN, which is modified by some distribution of absorption to produce the observed XRB, source counts and source colours.

Of the various models for the soft XLF (\eg \citealt{boyle94}, \citealt{page97}, \citealt{jones97}, \citealt{miyaji00}), we have chosen to use the Luminosity Dependent Density Evolution (LDDE1) XLF model of \citet{miyaji00}.
This was primarily because it is based on a large sample of AGN, and its model parameters have been determined for the currently preferred lambda-dominated cosmology.
The sample used to fit this XLF model contains a mixture of AGN both with, and without, broad lines, suggesting that it contains a subset of absorbed AGN. 
We adopt the best fitting parameter values presented in Table 3 of \citealt{miyaji00}, and where appropriate, have corrected for the $H_0 = 70$ km s$^{-1}$ Mpc$^{-1}$ used in this study. 
We integrate the XLF over the range $41 <$ log$L_{0.5-2} < 48$, $0.015 < z < 5$ to calculate the total number of AGN expected in the field.
A 1D cumulative probability distribution is generated by integrating the 2D XLF via an arbitrary path in $z$, $L_{0.5-2}$ space.
It is then possible to build a list of AGN which are randomly distributed in $z$, $L_{0.5-2}$ according to the model XLF.  
Each of these AGN are assigned a random value of $N_H$ according to the $f(N_H)$ model being tested, and a spectral slope taken from $g(\Gamma)$. 
The absolute normalisation of the XLF is iteratively adjusted, so that the simulated fields contain the same source counts as the \thirt sample at $S_{0.5-2} = 2 \times 10^{-15}$ \cgs.

\subsection[]{X-ray colours from AGN spectral templates}
We determine the X-ray colours of the simulated AGN by using a simple absorbed power-law (APL) model, which also includes a correction for the small Galactic absorbing column ($N_H \sim 8 \times 10^{19} $cm$^{-2}$) found in the \thirt field.
In order to compare simulated images with the observations, we must convert from the simulated AGN parameters to multi-band EPIC count rates.
We use the spectral fitting package \textsc{XSPEC} to generate fake spectra, incorporating both the instrumental response (for the MOS1, MOS2 and pn cameras), and the AGN parameters $z$, $N_H$ and $\Gamma$.
These spectra are summed over the appropriate energy bands, to derive the relevant conversion factors.
The cost in processing time would be prohibitive if we were to individually recalculate these conversion factors for each of the thousands of simulated AGN.
Hence, we have built lookup tables of conversion ratios, which finely sample $(z,N_H,\Gamma)$ parameter space, covering the range $0.01 < z < 5$, $19 <$ log$N_H < 25$, $1.2 < \Gamma < 2.6$. 
The conversion ratios are calculated for a single luminosity, but then scaled according to the luminosity of each simulated AGN.
These tables are used to convert rapidly from any set of simulated AGN parameters $L_X,z,N_H,\Gamma$ to count rates, for each EPIC detector, and energy band.

We have also examined the effect of including a small reflection component in the spectral model.
This has the net effect of hardening the spectrum at higher energies, making simulated AGN slightly more detectable above 5~keV.
We use the \textsc{PEXRAV} model of \citet{magdziarz95}, with the reflecting material covering $\pi$ steradians, a viewing angle of $30\deg$, and solar abundances.
We call this the APL+R spectral model.

It is beyond the scope of this study to include more complex AGN spectral features, such as $FeK$ lines, or scattered soft X-ray emission.
We expect the effect of these features on AGN colours to be small relative to the effects of continuum obscuration.
However, some of our results suggest that a number of detected sources in the \thirt field have an additional soft component, as discussed later.
We expect to be able to detect very few (if any) very heavily absorbed AGN having log$N_H > 25$, and so have not included such objects in the simulated populations.
In fact, the simulations show us that we expect AGN having log$N_H > 24$ to account for only $\sim 1\%$ of the detections in the \thirt sample.
Therefore, any additional attenuation due to Compton scattering within the dusty torus is ignored, since it has little effect for AGN with log$N_H < 24$.

\subsection[]{Imaging characteristics}
The simulation method incorporates the effects of the EPIC response function, effective area, point spread function (PSF), vignetting, and background to produce multi-band images.
We use the energy and off-axis angle dependent ``MEDIUM'' accuracy PSF model, taken from the \xmm calibration library.
This PSF model has been measured to be accurate to better than $\sim3\%$ at $1.5$keV \citep{gondoin00}.
The effective exposure time and vignetting are calculated from the SAS generated exposure maps for the \thirt field.
A synthetic background is added to the simulated images to reproduce the level observed in the observations.
The correct level of this additional background was determined through an iterative process to account for the contribution from the faint unresolved simulated sources.

\subsection[]{Source detection process applied to the simulated images}
We use the simultaneous, multi-band source detection process on the combined simulated MOS1+MOS2+pn images in the same fashion as described for the \thirt data \citep{loaring05}.
However, only one iteration of the background determination process is carried out, in order to conserve computation time.
We have searched for sources over the entire useful field of view of the combined EPIC detectors, giving a total sky area of 0.185 deg$^2$.

\section{Capabilities of the \thirt survey}
\label{section_capabilities}
The inherent capabilities and limitations of the \thirt survey data can be precisely evaluated using our simulation method.
In this section we refer to sources found in the simulated images by \textsc{EMLDETECT} as ``output'' sources.
We have employed a simple algorithm, that for each output source, associates an input source, in order that the colours of the output sources can be related to the input parameters ($z$, $L_X$, $N_H$, and $\Gamma$).
For the majority of output sources, there is a single nearby input candidate, which we consider to be the progenitor.
However, in some cases, an output source can have several nearby input candidate sources.
This problem is exacerbated at high off-axis angles because of the degradation of the PSF (and consequently, the precision of positions reported by \textsc{EMLDETECT}).
Therefore, we have employed a simple algorithm that matches each output source to the brightest input source within a small radius, $d$ of the detected position.
We make $d$ dependent on off-axis angle by setting it to 5\arcsec, 8\arcsec and 10\arcsec \ for off-axis angles $<$9\arcmin, 9\arcmin-12\arcmin, and $>$12\arcmin \ respectively.  
Any output sources with no input candidates within $d$ are almost certainly caused by Poissonian background fluctuations, and so are not considered in this section.
However, we expect a small number of the detections in the \thirt sample to be caused by this phenomenon, and hence unrelated to any real X-ray source.
Therefore, in section \ref{section_results}, we do include those output simulated sources having no input candidates when comparing the simulated AGN populations to the sample.

\subsection{Selection function}
\begin{figure}
\includegraphics[width=80mm]{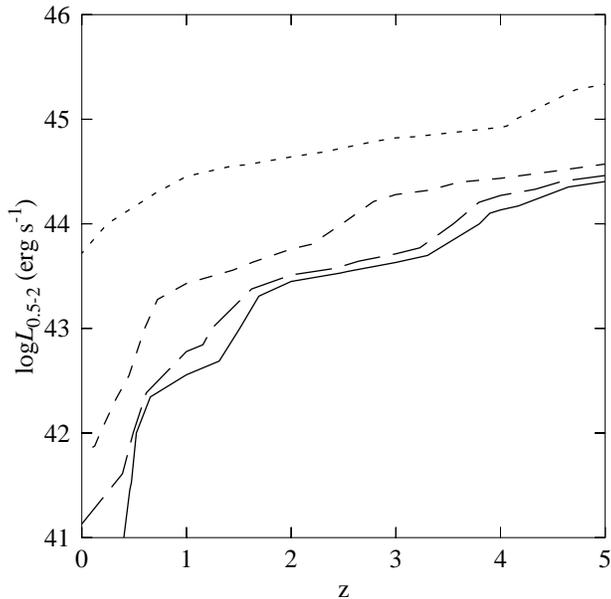}
\caption{The 50\% detection limits for simulated sources as a function of $z$ and $L_{0.5-2}$ for the absorbed power-law spectral model. The results for four ranges of absorption are shown: $0<$ log $N_H< 21.5$ (solid), $21.5<$ log$N_H<22.5$ (long dash), $22.5 <$ log$N_H < 23.5$ (short dash), and $23.5 <$ log$N_H < 25.0$ (dotted). }  
\label{det_frac_zL}
\end{figure}

To determine the selection function of the simulated AGN, we evaluate the fraction of matched output/input sources, as a function of the input parameters $z,L_X,N_H$, and $\Gamma$.
Fig. \ref{det_frac_zL} shows the 50\% completeness limit, as a function of redshift and luminosity, for several different levels of absorption.
The contours show the loci in $z-L_{0.5-2}$ parameter space, at which half of the input sources are detected.
There is a clear reduction in detection probability for AGN having absorbing columns above $10^{23.5}$ cm$^{-2}$, and this effect is more marked at low-$z$.
The plot shows that the \thirt survey is able to detect the majority of moderately luminous AGN (log$L_{0.5-2} \ge 44$), with moderate-absorption ($21.5 < $log$N_H < 22.5$) up to $z\approx 3.5$.

\begin{figure}
\includegraphics[angle=270,width=80mm]{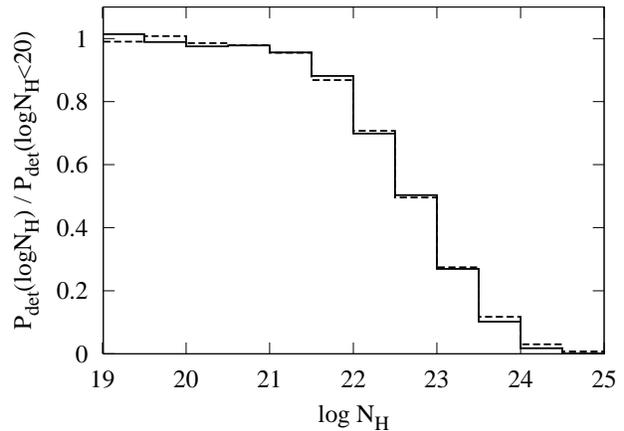}
\caption{The fraction of simulated input sources that are matched with an output source ($P_{det}$), as a function of $N_H$, normalised to the output/input fraction for log$N_H < 20$ sources. Results are shown for the APL spectral model (solid line), and the APL+R model (dashed line). The plot is compiled from the simulations for the $\beta =8$ $f(N_H)$ model.}
\label{nh_det_frac}
\end{figure}

Fig. \ref{nh_det_frac} shows the fraction, as a function of absorption, of all input sources that are matched to output sources in the simulated images.
This highlights the small differences in detection probability between the two spectral models.
The addition of a reflection component in the AGN spectra has a rather small effect on the detectability of simulated sources.

\begin{figure}
\hspace{-5mm}
\includegraphics[angle=270,width=90mm]{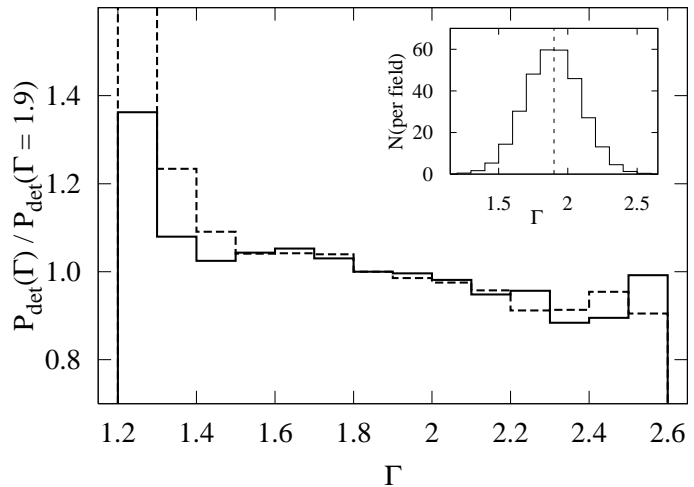}
\caption{The fraction of simulated input sources that are matched with an output source ($P_{det}$), as a function of $\Gamma$, normalised to the output/input fraction for $\Gamma = 1.9$ sources. Results are shown for the APL spectral model (solid line), and the APL+R model (dashed line). The plot is compiled from the simulations for the $\beta =8$ $f(N_H)$ model. The inset shows the number of output sources, per simulated field, as a function of spectral slope. }
\label{gamma_det_frac}
\end{figure}
The dependence of the selection function on $\Gamma$ can be seen in fig. \ref{gamma_det_frac}, which shows the fraction of simulated input sources with output counterparts, as a function of spectral slope.
It can be seen that the spectral slope of an AGN has a small but measurable bearing on its probability of detection.
A strong increase in detection probability is seen for very hard sources ($\Gamma < 1.4$), however, the inset histogram shows that very few of these objects are predicted by the $g(\Gamma)$ model. 
We have used the 0.5--2~keV de-absorbed flux to normalise the model spectra, and so the hard-slope AGN have a relatively high countrate above 2~keV, and are more likely to be detected.
This effect is larger for moderate to heavily absorbed sources, since they are primarily detected at these harder energies.
The impact on the overall selection function is largest for the $f(N_H)$ models containing the largest fraction of absorbed sources, i.e. the $R=8$, $R=R(z)$ models.

\subsection{Sensitivity to X-ray colours}
\begin{figure*}
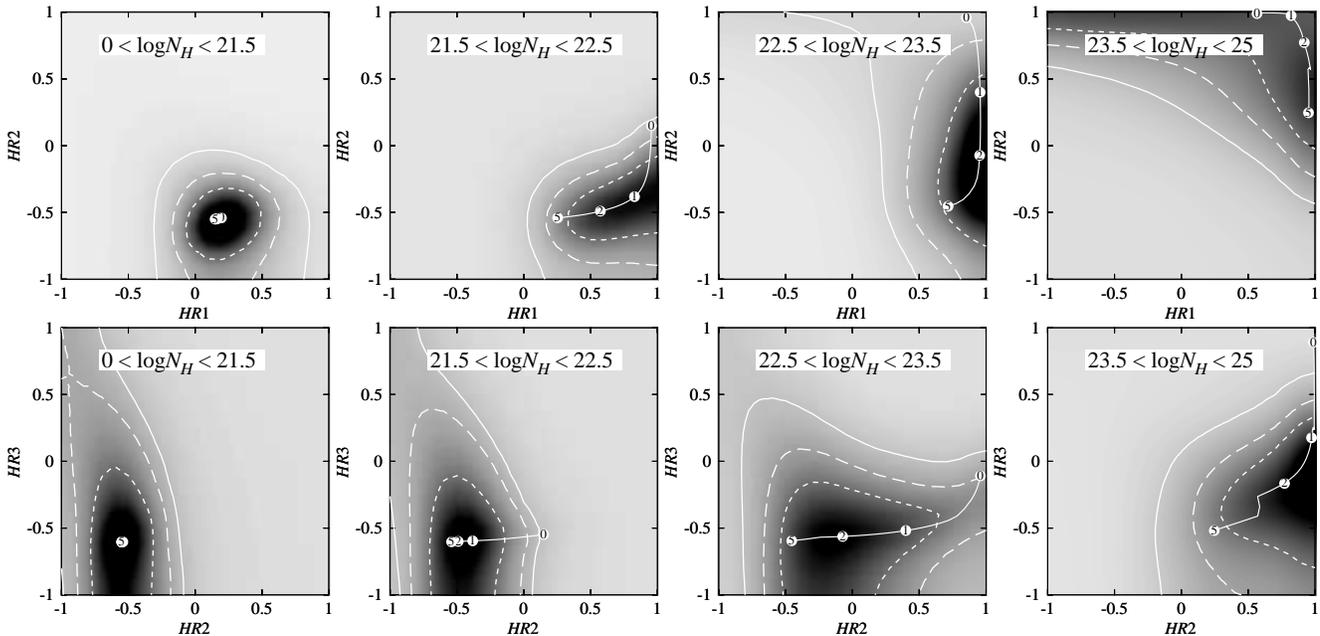

\includegraphics[width=43mm]{fig5_1.epsi}
\includegraphics[width=43mm]{fig5_2.epsi}
\includegraphics[width=43mm]{fig5_3.epsi}
\includegraphics[width=43mm]{fig5_4.epsi}
\includegraphics[width=43mm]{fig5_5.epsi}
\includegraphics[width=43mm]{fig5_6.epsi}
\includegraphics[width=43mm]{fig5_7.epsi}
\includegraphics[width=43mm]{fig5_8.epsi}

\caption{Colour-colour distributions of simulated sources for the APL spectral model, showing $HR1$ vs $HR2$ (upper row) and $HR2$ vs $HR3$ (lower row). The panels show the colours produced by simulated sources grouped into bins according to their intrinsic absorption ($0 <$ log$N_H < 21.5$, $21.5 < $ log$N_H < 22.5$, $22.5 <$ log$N_H < 23.5$, and $23.5 <$ log$N_H < 25.0$). 
The levels of the contours are set such that they include $50\%$ (short dash), $75\%$ (long dash), and $90\%$ (solid line) of the sources. The contribution of each simulated source to the greyscale map was represented by an ellipical Gaussian centered on the measured position in colour-colour space, and having widths equal to the corresponding $\sigma_{HR}$. 
We also show the locus of expected colours for an AGN with an APL spectrum, $N_H$ in the logarithmic center of the interval, $\Gamma = 1.9$, and with $0 < z < 5$ (numbered points indicate $z$).}
\label{colour_nh_ranges}
\end{figure*}

Constraints on $f(N_H)$ models can be made from analysis of X-ray colour (i.e hardness ratio) distributions.
For example, \citet{perola04} compared the $N_H$ of \xmm sources determined from full spectral fits, with the $N_H$ estimated using a hardness ratio method (over the 0.5--10 keV range), and showed that they were in good agreement for log$N_H > 22$. 
For this study we define the hardness ratios $HR1 = (R_{0.5-2}-R_{0.2-0.5})/(R_{0.5-2}+R_{0.2-0.5})$, $HR2=(R_{2-5}-R_{0.5-2})/(R_{2-5}+R_{0.5-2})$ and $HR3=(R_{5-10}-R_{2-5})/(R_{5-10}+R_{2-5})$, where $R_{E_{min}-E_{max}}$ is the source count rate, corrected for vignetting, in the given energy band.
The corresponding measurement errors are denoted by $\sigma_{HR1}$, $\sigma_{HR2}$, $\sigma_{HR3}$ respectively. 
The count rates, hardness ratios, and errors are computed within \textsc{EMLDETECT} using the combined dataset from the MOS1, MOS2 and pn cameras.
If any hardness ratio measurement is undetermined (zero countrates in two energy bands), we set it to $0.0 \pm 1.0$.

The dependence of $HR1$, $HR2$ and $HR3$ on absorption is illustrated in fig. \ref{colour_nh_ranges}, which shows the measured colour-colour distributions of ``output'' simulated sources grouped into a number of $N_H$ bins.
For each $N_H$ bin, we have over-plotted the ``perfect'' $z$-track in colour-colour space for an AGN with mid-bin absorption, $\Gamma = 1.9$, and $0 < z < 5$.
Both the width of the $N_H$ bins, and the range of $\Gamma$ in the simulated sources, act to distribute sources about this track.
The relative density of the distribution along the track is mostly determined by the evolution of the XLF, which peaks above $z=1.5$.
In addition, a significant amount of scatter is caused by measurement uncertainties within the source detection process, particularly for the faintest sources. 

We see from the left-most upper panel of fig. \ref{colour_nh_ranges}, that the colour distribution of output simulated sources with log$N_H < 21.5$ is compact, and approximately centered on ($HR1,HR2$) = (0.2,-0.5).
The study of \xmm sources in the Lockman hole by \citet{mainieri02}, found that the vast majority of AGN in this part of hardness ratio space had broad line optical counterparts and, at most, weak absorption (log$N_H<21.5$) in their X-ray spectra.
In contrast most of the identified AGN having hard X-ray colours had narrow line optical counterparts, although only a small fraction of the hard sources had optical identifications.
Examination of the three upper right panels reveals that the moderately to heavily absorbed sources (log$N_H>21.5$), occupy a measurably different region of $HR1$,$HR2$ space compared to their less absorbed counterparts.
In particular, $HR1$ is sensitive to absorption in the range $21.5<$log$N_H<23.5$, and $HR2$ to absorption above log$N_H=22.5$.
In the study of \citet{georga04}, the hardness ratio between the 0.5--2 and 2--8 keV bands did not appear to separate the broad and narrow line AGN; however, relatively few of the harder AGN in this sample had spectroscopic identifications.
\citet{dellaceca04} showed that the majority of AGN with broad line counterparts fall in the range $-0.75 < HR2 < -0.35$, consistent with the location of the low absorption AGN (log$N_H<21.5$) produced by our simulations.

As we would expect, the majority of the simulated faint unabsorbed AGN do not have good measurements of $HR3$.
These sources have noise dominated countrate measurements above 2~keV, and hence have $HR3$ measurements randomly scattered in the interval $[-1,1]$.
Of the simulated AGN having log$N_H>23.5$, it is only the most luminous ($L_{0.5-2}>10^{44}$ erg~s$^{-1}$), that are detectable in our survey, as shown in figs. \ref{det_frac_zL} and \ref{nh_det_frac}.
The bottom right hand panel illustrates that $HR3$ is sensitive to absorption above log$N_H = 23.5$ for all but the highest redshift AGN. 
Hard band X-ray count rates were well determined for sources in the bright sample of \citet{caccianiga04}, and of the four objects with a higher count rate in the 4.5--7.5 keV band than in the 2--4.5 keV band, three were associated with narrow line optical counterparts, and one with a Seyfert 1.9 galaxy.

\begin{figure*}
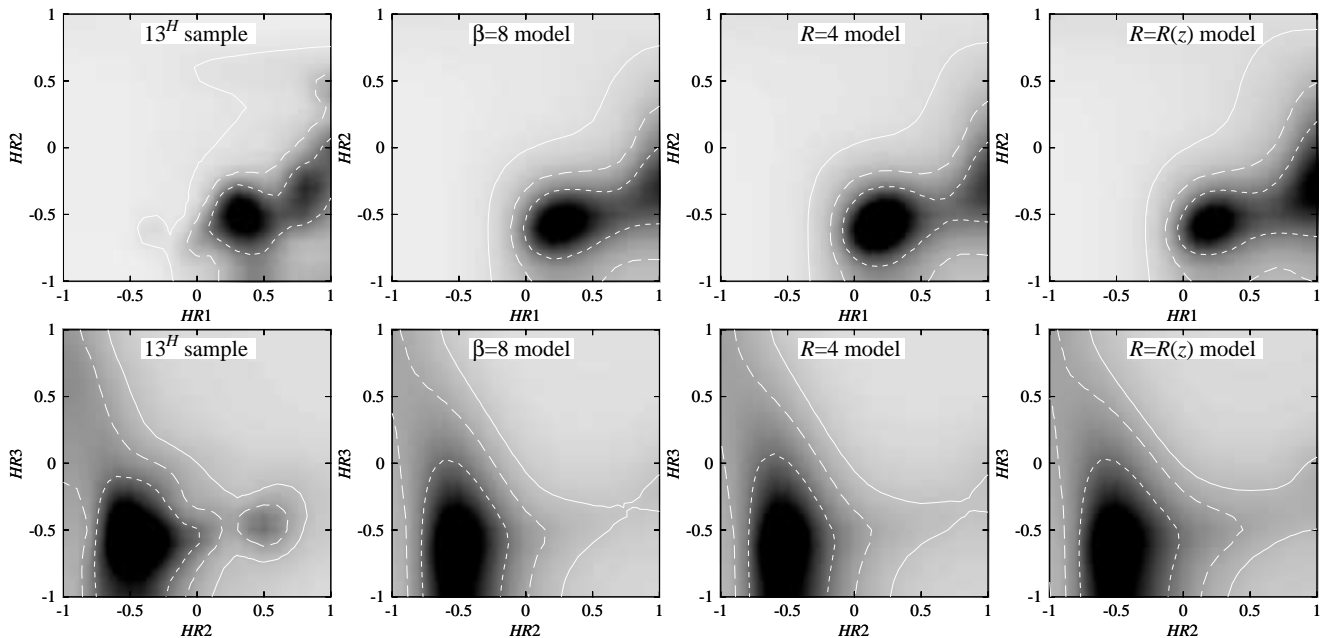

\includegraphics[width=43mm]{fig6_1.epsi}
\includegraphics[width=43mm]{fig6_2.epsi}
\includegraphics[width=43mm]{fig6_3.epsi}
\includegraphics[width=43mm]{fig6_4.epsi}
\includegraphics[width=43mm]{fig6_5.epsi}
\includegraphics[width=43mm]{fig6_6.epsi}
\includegraphics[width=43mm]{fig6_7.epsi}
\includegraphics[width=43mm]{fig6_8.epsi}

\caption{Colour-colour distributions produced by different $f(N_H)$ models compared to that seen in the sample data. The panels show the \thirt data (top left), and then the results for three of the simulated $f(N_H)$ models (using the absorbed power-law spectral model). The levels of the contours are set such that they include $50\%$ (short dash), $75\%$ (long dash), and $90\%$ (solid line) of the smoothed source distribution, and were generated in the same way as for fig. \ref{colour_nh_ranges}.} 
\label{colours_sims_models}
\end{figure*}

Figure \ref{colours_sims_models} shows the $HR1$ {\it vs.} $HR2$ (upper row) and $HR2$ {\it vs.} $HR3$ (lower row) distributions produced by three of the $f(N_H)$ models. 
The most immediately noticeable difference between the plots, is the fraction of sources that appear to the right of $HR1 = 0.6$ for the various $f(N_H)$ models.

\section{Results}
\label{section_results}

\subsection{Colour distribution of the \thirt sample}
\label{section_colour_distribution}
\begin{figure}
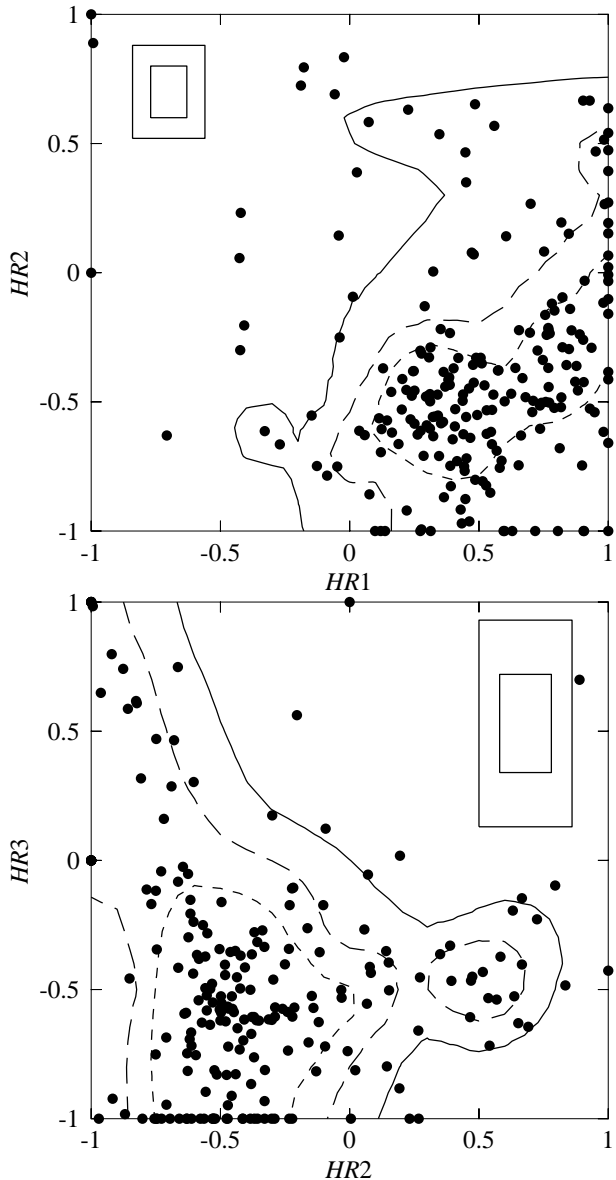

\includegraphics[width=80mm]{fig7_1.epsi}
\includegraphics[width=80mm]{fig7_2.epsi}
\caption{X-ray colour-colour distributions found in the \thirt sample. The levels of the contours are set such that they include $50\%$ (short dash), $75\%$ (long dash), and $90\%$ (solid line) of the smoothed source distribution, and were generated in the same way as for fig. \ref{colour_nh_ranges}. Typical sizes of $\sigma_{HR1}$, $\sigma_{HR2}$, and $\sigma_{HR3}$ are shown with boxes, for sources having ``two-band'' fluxes of $10^{-14.5}$ and $10^{-14}$ \cgs, where the ``two-band'' flux is the flux measured over the two energy bands used to calculate the hardness ratio.}
\label{colours_13hr}
\end{figure}

The two left-most panels of fig. \ref{colours_sims_models} show the colour-colour distributions of the \thirt sample, with grey-scale and contours generated in the same way as for fig \ref{colour_nh_ranges}.
Fig. \ref{colours_13hr} shows the same contours, but with the individual data points overlaid.
Figs. \ref{colours_sims_models} and \ref{colours_13hr} show that there is a strong concentration of sources in the $(HR1,HR2,HR3)$ = (0.4,-0.5,-0.5) region, slightly harder in $HR1$ than the nominal position of an unabsorbed AGN with log$N_H < 21$, $\Gamma = 1.9$.
A large number of sources have much harder values of $HR1$ and $HR2$ than the nominal unabsorbed position, indicating that strong absorption is present in a significant fraction of the population.
However, the majority of the sources in the $HR2~<~0$, $HR3~>~-0.3$ region are actually faint soft sources having large $HR3$ measurement uncertainties. 
The bimodality apparent either side of $HR1 = 0.6$ is probably due to the fast increase in $HR1$ over the range $21.5~<~$~log$N_H~<~22.5$, which limits the number of sources in this region.
A similarly sparse region occurs at $HR2 \sim 0.25$ and again, this is probably related to the fast increase in $HR2$ over the range $23~<~$~log$N_H<~24$.

\subsection{Reproducing the observed source counts}
\label{section_source_counts}
\begin{figure}
\hspace{-5mm}
\includegraphics[angle=270,width=90mm]{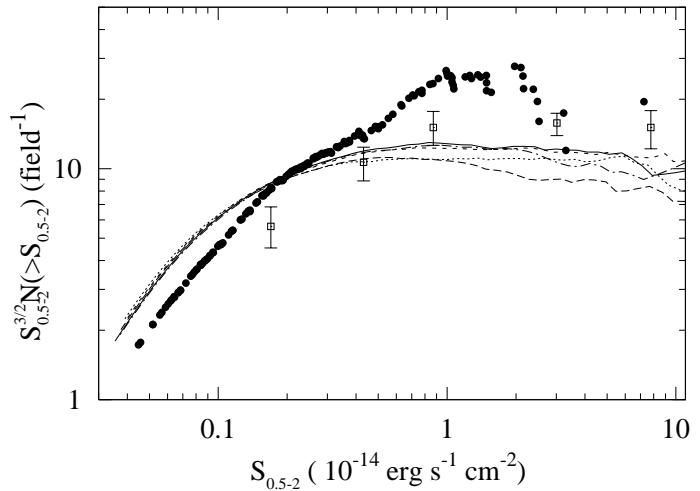}
\caption{The $N(> \!\! S_{0.5-2})$ curves measured in the \thirt sample (filled circles), compared to those produced by the $\beta = 2$ (solid line), $\beta =8$ (long dash), $R=4$ (short dash), $R=R(z)$ (dotted), and $R=R(L_X)$ (dot-dash), $f(N_H)$ models. These results are for the absorbed power-law spectral model and are normalised to the Euclidean-slope. The equivalent points for the sample of \citealt{miyaji00} (taken from fig. 6), are also shown (open symbols with errorbars), and have been normalised assuming a sky area of 0.185 deg$^2$. }
\label{source_counts_soft}
\end{figure}

We have compared the 0.5--2 keV band integral source counts, $N(>S_{0.5-2})$, measured in the \thirt sample with those produced by the simulated model distributions. 
We make no correction for sky coverage, since the \thirt sample and the simulated fields have an identical survey-depth/sky area relation.
We find a large disparity between the \thirt and simulated fields, especially around the knee of the observed $N(> S_{0.5-2})$ at $\sim 10^{-14}$ \cgs, as can be seen in fig. \ref{source_counts_soft}.
Each of the $f(N_H)$ models produced similar $N(>S_{0.5-2})$ curves, especially at faint fluxes, where the statistical errors are better.
Thus we deduce that the data-model disparity is primarily caused by differences between the data and the model XLF (and/or its evolution).
We discuss this later.	
The primary purpose of this study is to compare the $f(N_H)$ models, so it is important that we minimise the effect on the statistical analysis caused by the disparity between the data and XLF/evolution model.
Therefore, we have examined the X-ray colour distribution of sources, rather than the distributions of their absolute fluxes.
We expect the colour-colour distributions to be more sensitive to $f(N_H)$ than to the XLF, because a small change of the position of an absorbed AGN in the $z$-$L_X$ plane, has a strong effect on its overall brightness, but only a small effect on its X-ray colours.
For example, if the peak of AGN space density is actually at $z=1.3$, (rather than at $z=1.6$ as predicted by the XLF/evolution model), then the resulting change in hardness ratios for an AGN, having an absorbed power-law spectrum, log$N_H = 22$, at this peak redshift, would be $\Delta HR1(\Delta HR2)$ = +0.07(+0.03). 
However, an increase of 0.5 dex in the absorption of the same AGN, would result in $\Delta HR1(\Delta HR2)$ = +0.23(+0.17).
Therefore, a colour analysis of the $f(N_H)$ models is more strongly dependent on the tested $f(N_H)$, than on differences between the data and XLF/evolution model.

\subsection{Statistical comparison of colour distributions in the data and the models}
\label{section_colour_stats}
We have used the Kolmogorov-Smirnov test (KS), to determine how well the simulated data reproduce the X-ray colour distribution measured in the \thirt sample.
The KS test has the advantage that it requires no rebinning of data, utilising the full information content of the data set.
However, it does not take into account the relative errors on data points, meaning that low signal to noise measurements can, to some extent, ``wash out'' the signal from the more precise measurements. 
A three dimensional extension of the KS test (3D-KS), as devised by \citet{fasano87}, was used to compare the sample with the simulation results in the full (HR1,HR2,HR3) variable space.
In order to examine more closely how the models reproduce the sample distribution, we have carried out one dimensional KS tests separately on $HR1$, $HR2$ and $HR3$.

The conversion from the 3D-KS test statistic, $D_{3D-KS}$, to the probability that two samples were taken from the same underlying population, $P_{3D-KS}$, is strongly dependent on the number of sources in the tested samples, and the degree of correlation between the tested variables. 
\citet{fasano87} numerically generated lookup tables to allow this conversion at a number of confidence levels, for a range of sample sizes, and values of the correlation parameter.
However, these tables give only a relatively small number of conversion values, at discrete confidence levels, sample sizes, and values of the correlation parameter.
Therefore, we have run a set of simulations, the results of which permit conversion from $D_{3D-KS}$ directly into $P_{3D-KS}$ conversion using the precise sample sizes and correlations seen in the \thirt sample.
We calculated the three-dimensional probability density map (3D-PDM), of the \thirt sample in (HR1,HR2,HR3) space.
The contribution from each source to the 3D-PDM is calculated from a 3D-Gaussian that has widths equivalent to $\sigma_{HR1}$,$\sigma_{HR2}$, and $\sigma_{HR3}$.
The normalisation of the 3D-Gaussian is set such that the total contribution of each source is unity.
This 3D-PDM is used to generate pairs of random populations, having 217 and 25000 members respectively, for which $D_{3D-KS}$ is calculated.
The latter step is repeated for 100000 iterations.
The equivalent probability for any particular value of the 3D-KS statistic, is equal to the fraction of these iterations having $D_{3D-KS}$ greater than this value.
The absolute lower limit at which we can evaluate the probability is given by the reciprocal of the number of simulation iterations, i.e. 0.001\%, although the errors are large at this level. 
This limit is determined by the processing time available.
Table \ref{KS_results} shows $P_{3D-KS}$, for the eight $f(N_H)$ models, and for both of the tested spectral models.
In order to determine where the biggest differences arise between the data and models, we have calculated the KS probabilities ($P_{KS}$), separately for each of $HR1$, $HR2$ and $HR3$, the results of which are also shown in table \ref{KS_results}.

\begin{table*}
\begin{minipage}{180mm}
 \caption{ 3D-KS and KS test probabilities, calculated by comparing the distributions of $HR1$, $HR2$ and $HR3$ produced by the eight tested $f(N_H)$ models with that found in the \thirt sample.  Results are shown for both the absorbed power-law (APL), and absorbed power-law with reflection (APL+R), spectral models.}
 \label{KS_results}
 \begin{tabular}{@{}lrrrrrrrr}
  \hline
 		      & \multicolumn{2}{c}{$P_{3D-KS}(HR1,HR2,HR3)$} & \multicolumn{2}{c}{$P_{KS}(HR1)$}      &  \multicolumn{2}{c}{$P_{KS}(HR2)$}     & \multicolumn{2}{c}{$P_{KS}(HR3)$}\\
  \hline
  $N_H$ Model         & APL                 &   APL+R                & APL                 &   APL+R             & APL                  & APL+R              & APL                 & APL+R   \\
  \hline
  \textbf{$R=0$}      & $<1 \times 10^{-5}$ & $<1 \times 10^{-5}$    & $<1\times 10^{-10}$ & $<1\times 10^{-10}$ & $<1\times 10^{-10}$ & $<1\times 10^{-10}$ & $2.6\times 10^{-4}$ & $0.0027$ \\
  \textbf{$\beta=2$}  & $<1 \times 10^{-5}$ & $1  \times 10^{-5}$    & $3.6\times 10^{-7}$ & $3.1\times 10^{-6}$ & $1.4\times 10^{-4}$ & $0.0016$            & $0.0052$            & $0.034$ \\
  \textbf{$\beta=5$}  & $0.0021$            & $0.0019$               & $2.5\times 10^{-4}$ & $3.9\times 10^{-4}$ & $0.0050$            & $0.057$             & $0.0078$            & $0.039$ \\
  \textbf{$\beta=8$}  & $0.0074$            & $0.058$                & $0.019$             & $0.041$             & $0.20$              & $0.77$              & $0.013$             & $0.094$  \\
  \textbf{$R=4$}      & $<1 \times 10^{-5}$ & $<1 \times 10^{-5}$    & $1.3\times 10^{-8}$ & $4.5\times 10^{-8}$ & $0.081$             & $0.37$              & $0.014$             & $0.043$ \\
  \textbf{$R=8$}      & $0.0014$            & $0.0033$               & $0.012$             & $0.015$             & $0.96$              & $0.62$              & $0.025$             & $0.12$  \\
  \textbf{$R=R(z)$}   & $3.4\times 10^{-4}$ & $0.0016$               & $0.0058$            & $0.0019$            & $0.88$              & $0.39$              & $0.028$             & $0.19$  \\
  \textbf{$R=R(L_X)$} & $<1 \times 10^{-5}$ & $1 \times 10^{-5}$     & $1.5\times 10^{-6}$ & $4.2\times 10^{-6}$ & $4.3\times 10^{-4}$ & $0.0033$            & $0.0038$            & $0.026$ \\
 \end{tabular}
\end{minipage}
\end{table*}

\section{Selecting absorbed sources}
\label{section_select_absorbed}
\begin{figure}
\includegraphics[angle=270,width=80mm]{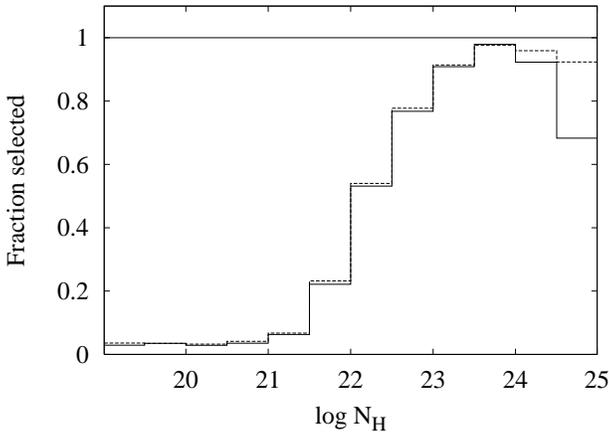}
\caption{The fraction of output simulated sources which are selected by the hardness ratio cut $HR1 - \sigma_{HR1} > 0.6$ OR $HR2 - \sigma_{HR2} > -0.3$, as a function of $N_H$ .The solid line shows the result for the absorbed power-law spectral model, and the dashed line shows the result with an addition of a reflection component. }
\label{NH_histo_hard}
\end{figure}

If we wish to examine just moderate to heavily absorbed AGN, then we need some X-ray colour selection criteria which will allow us to choose only this population.
Examination of the simulation results (see fig. \ref{colour_nh_ranges}), shows that a cut of $HR2 > -0.3$ will select the majority of the most heavily absorbed sources (log$N_H > 22.5$).
This $HR2$ cut is the same as that shown to discriminate efficiently between optical type-1 and type-2 AGN in the \xmm Bright Serendipitous Survey (XBSS) sample \citep{caccianiga04,dellaceca04}.
However, it should be cautioned that the XBSS definition of $HR2$ differs slightly from our own; they use the 2--4.5 keV energy band (rather than our 2--5 keV band), and report $HR2$ only for the MOS2 dataset.
AGN with absorption in the range $21.5~<~$log$N_H~<~22.5$ are included by adding the region with $HR1 > 0.6$.
To reduce the number of faint soft sources, having low signal-to-noise measurements, that are scattered into the ``hard'' sample, we require that a source satisfies the above conditions by more than $1\sigma$ to be included.
Fig. \ref{NH_histo_hard} demonstrates the effectiveness of such a cut in selecting only those sources with significant $N_H$. 
The slight dip in the selected fraction at log$N_H > 24$ is caused by the generally larger errors on $HR1/HR2$ for the most heavily absorbed sources. 
This evaluation of the effectiveness of the selection scheme assumes a simple absorbed power-law AGN spectral model. 
Spectral features such as an additional soft component, will serve to degrade this efficiency.
Because of the relatively poor average measurement accuracy for $HR3$, we have not used it to select absorbed sources.
These ``hard'' selection criteria, when applied to the \thirt sample, result in 86 hard sources ( 39\% of the total).
This value is consistent with the fraction ($34 \pm 9 \%$) of {\em optical} type-2 AGN identified in the 2--4.5~keV selected subset of the XBSS \citep{dellaceca04}.
The ``hard'' fraction for each of the $f(N_H)$ models are presented in table \ref{hard_fraction}. 
We see that the fractions of hard sources produced by the $\beta=8$, $R=8$, and $R=R(z)$ models are consistent within $3\sigma$ with the fraction seen in the \thirt sample.
By including a reflection component in the spectra, the hard fraction is increased by less than 2\% for all but the $R=R(L_X)$ model.

\section{Discussion}
\label{section_discussion}
\subsection{Reproducing the colours in the sample}
Table \ref{KS_results}, shows that, when we consider an absorbed power-law spectral model, none of the $f(N_H)$ models provide a good description of the X-ray colours in the \thirt sample, and are all strongly rejected by the 3D-KS test (with greater than 99\% confidence).
The $\beta = 8$ model provides the best fit with a probability of 0.8\%; although low, this value includes the effects of the disparity between the data and the XLF/evolution model.
However, the addition of a reflection component to the AGN spectra improves the $P_{3D-KS}$ for almost all of the $f(N_H)$ models.
The best fitting distribution is still the $\beta=8$ model, but with a much improved probability of 6\%.
The remainder of the $f(N_H)$ models are strongly rejected by the 3D-KS test, with greater than 99.5\% probability. 
This match between the $\beta=8$ model and the sample, is actually rather a good one, considering that the only tuned parameter is the overall normalisation of the XLF.
The large range of $P_{3D-KS}$ that is measured between the $f(N_H)$ models, demonstrates that our colour analysis technique is indeed a good probe of the underlying distribution of absorption in the sample.

The results of KS tests on individual hardness ratios reveal more clearly where the $f(N_H)$ models succeed or fail to reproduce the sample colours.
When $HR3$ is considered, the addition of a reflection component to the absorbed power-law spectral model improves the fit of all the tested $f(N_H)$ models.
However, there is not such a consistent improvement in the fits for $HR1$ and $HR2$.
The $HR2$ distributions produced by the $\beta=8$, $R=4$, $R=8$ and $R=R(z)$ models closely match the distribution seen in the \thirt sample, with KS probabilities greater than 30\% when a reflection component is included in the model spectra.
However, large differences between the models and the sample arise in the distributions of $HR1$ and $HR3$.
The tested $f(N_H)$ models over-produce the fraction of sources having very hard colours below 2 keV ($HR1 = 1$) relative to the \thirt sample. 
In the \thirt sample, the fraction of sources having $HR1 = 1$ is around 8\%; however, even in the best fitting $\beta=8$ model, the fraction is around 15\%.
We find that almost all (90\%) of the simulated sources having $HR1 = 1$ have log$N_H > 22$, and so have had virtually all their flux below 0.5~keV removed by absorption. 
However, these sources are distributed similarly to the rest of the population in $L_X$, $z$ and $\Gamma$. 
The relatively low $P_{KS}(HR1)$ for the $R=4$, $R=8$ and $R=R(z)$ models, can be partly attributed to their low number of lightly absorbed AGN ($20~<$~log$N_H~<~22$). 
Within this subset of models, the evolving $R=R(z)$ model is strongly favoured over the $R=4$ model, but is marginally less successful than the $R=8$ model.
However, the latter model is unphysical in that it contains a much greater ratio of absorbed to unabsorbed sources than is seen in the local universe.
When the $HR3$ distribution is considered, we find the models produce too large a fraction of simulated sources having $HR3 = -1$.
This is most probably caused by the over-abundance of very faint sources produced by the simulations, related to the XLF mismatch.
These sources are detected just above the flux limit in the softer bands, but have count rates which fall below the background level in the hardest band, and hence are measured to have $HR3 \approx -1$. 

The statistical analysis strongly rejects the $R=R(L_X)$ model, in agreement with the findings of a recent study by \citet{treister04}, which was based on deep multi-wavelength data in the GOODS fields.
These authors tested the $f(N_H)$ model of \citet{ueda03}, alongside a simpler $f(N_H)$, but found that the latter provided a much better description of the data.

By examining the subset of sources satisfying the ``hard'' selection criteria, we can compare the distributions of absorption above log$N_H = 22$ that are found in the \thirt sample with those predicted by the models.
We have carried out 3D-KS and KS tests on $HR1$, $HR2$, and $HR3$, as before, but only for the ``hard'' selected subsets of the \thirt sample and simulations.
The 3D-KS test rejects each of the $f(N_H)$ models with high confidence, (both with and without a reflection component included in the model spectra).
We have examined the individual $KS$ test results to determine the source of this large disparity.
We find that the $KS$ probabilities for the best fitting $\beta=8$ model, (with the absorbed power-law spectral model), are 0.0003, 0.76, and 0.12, for $HR1$, $HR2$ and $HR3$ respectively.
The equivalent probabilities when an additional reflection component is included in the model spectra are 0.0002, 0.77, and 0.29.
The $KS$ test probabilities do not vary greatly between the different $f(N_H)$ models (excepting the $R=0$ model).
The $HR2$ and $HR3$ distributions of all the $f(N_H)$ models (excepting the $R=0$ model) provide rather good matches to the $HR2$ and $HR3$ distributions found in the ``hard'' subset of the \thirt sample.
The differences between the ``hard'' subsets of the $f(N_H)$ models are small, due to the rapid decline in the selected fraction of ``input'' sources for high absorbing columns (see fig. \ref{nh_det_frac}).
This acts to diminish the importance of the differences between the $f(N_H)$ models above $N_H = 10^{22}$ cm$^{-2}$.
The addition of a reflection component to the spectral model improves the KS probability for $HR3$ by a factor of $\sim 2$.

We see that the mismatch between the $HR1$ distributions is much worse in the ``hard'' subset, compared to the sample as a whole.
This appears to be due to the overproduction of simulated sources having $HR1 = 1$, which is more pronounced in the ``hard'' sub-sample. 
The fraction of the ``hard'' sample with $HR1 = 1$ is 20\% for the \thirt field, but $\sim 40\%$ for the model populations.
The disparity could be explained if a number of the heavily absorbed AGN have an additional soft X-ray component in their spectra.
In order to reproduce the distribution of $HR1$, this phenomenon should occur in around 10--20\% of the heavily absorbed sources. 
A number of absorbed AGN with excess soft emission have been observed by other authors in samples of spectroscopically identified X-ray sources (e.g. \citealt{caccianiga04}, \citealt{page05}).
This excess component could be due to intense starbursts in the host galaxy, or to diffuse emission surrounding an AGN embedded in a galaxy cluster.
Alternatively, it could be scattered radiation from the central engine of the absorbed AGN.

\begin{table}
\caption{Fraction of ``hard'' sources, $h$, (satisfying $HR1 - \sigma_{HR1} > 0.6$ OR $HR2 - \sigma_{HR2} > -0.3$), produced by each simulated $f(N_H)$ model. The corresponding fraction seen in the \thirt sample is 0.39 (86/217). Results are shown for an absorbed power-law spectral both with (APL+R) and without (APL) a reflection component. The standard deviation of the hard fraction $\sigma_h$, over the 100 simulation repetitions is also shown.}
\label{hard_fraction}
\begin{tabular}{@{}lrrrr}
 \hline
                     &   \multicolumn{2}{c}{APL} & \multicolumn{2}{c}{APL+R} \\
$f(N_H)$             & $h  $      & $\sigma_h$   & $h  $      & $\sigma_h$ \\
 \hline
\textbf{$R=0$}       &    0.026   & 0.010        & 0.029      & 0.010 \\ 
\textbf{$\beta=2$}   &    0.226   & 0.024        & 0.239      & 0.026 \\ 
\textbf{$\beta=5$}   &    0.278   & 0.025        & 0.288      & 0.024 \\ 
\textbf{$\beta=8$}   &    0.332   & 0.025        & 0.347      & 0.029 \\ 
\textbf{$R=4$}       &    0.300   & 0.023        & 0.313      & 0.029 \\ 
\textbf{$R=8$}       &    0.399   & 0.024        & 0.413      & 0.028 \\ 
\textbf{$R=R(z)$}    &    0.420   & 0.022        & 0.436      & 0.028 \\ 
\textbf{$R=R(L_X)$}  &    0.235   & 0.024        & 0.240      & 0.026 \\ 
\end{tabular}			  
\end{table}

\subsection{Implications for torus models}

For the simplest toy model of a torus with uniformly density, and a typical opening angle, $\theta_o$, the fraction of AGN that are heavily absorbed is approximately $cos(\theta_o)$.
So, if we use the size of the ``hard'' fraction of the \thirt sample as a measure of the number of absorbed AGN, we can infer a rather wide opening angle of $\theta_o \sim 67\degr$. 
However, this estimate does not take into account the effect of the drop in the selection function toward high $N_H$, and can only be seen as an upper limit on $\theta_o$.
We estimate the relative selection function for hard sources by counting the fraction of simulated ``hard'' input sources that have output counterparts relative to that for all input sources. 
Applying this correction to the \thirt sample, we predict an {\em intrinsic} ``hard'' fraction of $\sim 0.8$, implying an opening angle of  $\theta_o \sim 37\degr$.
If in our correction for the relative selection function, we exclude those sources with absorbing column above log$N_H = 24$, where our sample constrains the models only weakly, then we find $\theta_o \sol 52\degr$.
We are also able to examine the range of torus parameters that would best match the $f(N_H)$ models.
For the best fitting $\beta = 8$ model, where the fraction of input sources with log$N_H > 22$ is $\sim 0.75$, the predicted opening angle is $\theta_o \sim 42\degr$.

As fig. \ref{colour_nh_ranges} shows, $HR1$ and $HR2$ are sensitive to the shape of the distribution over a wide range of $N_H$, particularly for intermediately absorbed sources.
We have seen that the $\beta = 8$ is strongly favoured over the $R=4$ $f(N_H)$ model (see table \ref{KS_results}). 
These two models are very similar in the range $22 < $~log$N_H < 24$, contain similar numbers of unabsorbed AGN (log $N_H < 21$), and produce comparable numbers of ``hard'' sources.
Therefore, the difference must lie primarily in the $21 <$~log$N_H < 22$ range, in which the $\beta = 8$ model contains many more AGN.
A major problem with the uniformly dense torus model is that it predicts that nearly all AGN will be either heavily absorbed or completely unabsorbed. 
However, more complex models, incorporating a wide distribution of torus densities, predict larger numbers of intermediately absorbed AGN.
For example, a model in which the density falls off exponentially with angle away from the plane of the torus, predicts a much flatter $f(N_H)$ (e.g. \citealt{treister04}).
It is possible, with some tuning of such a model's parameters, to approximately match the best fitting $\beta = 8$ distribution.

Since absorption in the $21 <$~log$N_H < 22$ range has only a significant effect on $HR1$, it would not have been detectable in the colour distributions if the $0.2-0.5$ keV band had not been considered.
A number of studies of absorption in faint AGN have based their estimates of $N_H$ on hardness ratios between the $0.5-2$ and $2-10$ keV bands, and therefore may have underestimated the number of intermediately absorbed AGN (e.g. \citealt{ueda03},\citealt{treister04}). 

A better determination of mean torus properties will be possible when the \thirt field is covered by \spitzer, and we are able to correlate X-ray colours with mid/far-IR data.

\subsection{Source count disparity}
Each of the simulated $f(N_H)$ models produced similar 0.5--2.0 keV source count-flux relations, $N(>S_{0.5-2})$.
However, these are seen to reproduce poorly the $N(>S_{0.5-2})$ relation observed in the \thirt sample (see fig. \ref{source_counts_soft}).
The models under-produce the $N(>S_{0.5-2})$ above the normalisation flux ($2 \times 10^{-15}$ \cgs), and over-produce the $N(>S_{0.5-2})$ below this flux (see fig. \ref{source_counts_soft}).
In fact, at $10^{-14}$ \cgs the models under-produce the source counts seen in the \thirt sample by a factor of about two.
This disparity is seen to a similar degree in each of the $f(N_H)$ models, suggesting that it is related to the difference between the data and XLF/evolution model.
The $N(>S_{0.5-2})$ of the \citet{miyaji00} sample, is also shown in fig. \ref{source_counts_soft}, plotted assuming our field has a uniform sky area of 0.185 deg$^2$. 
This illustrates that in the flux range $10^{-14} - 10^{-13}$ \cgs, the LDDE1 XLF also under-produces the source counts of the sample from which it was derived.
The shape of the $N(>S_{0.5-2})$ relation of the \citet{miyaji00} sample is closer to that seen in the \thirt sample than to the models. 
The faintest AGN in the \citet{miyaji00} sample are from the deepest \rosat observations of the Lockman Hole field, where the flux limit of the data was $\sim 2 \times 10^{-15}$ \cgs.
Our significantly deeper flux limit means that we are using part of $L-z$ space outside that constrained by the sample of \citet{miyaji00}.
A previous comparison of source counts from \rosat observations in the Lockman Hole and \thirt fields, revealed a $\sim 10-20 \%$ over-abundance near $S_{0.5-2} = 10^{-14}$ \cgs in the \thirt field with respect to the Lockman Hole \citep{mchardy98}. 
In addition, \citet{loaring05} found that the \thirt field is slightly over-dense in the 0.5--2 keV band with respect to both of the Chandra deep fields.
Therefore we conclude that the differences between model and sample are caused by a combination of these factors.
In particular, our extrapolation of the LDDE1 XLF/evolution model to faint fluxes, suggests that this complex scheme requires some revision.

\subsection{High-$z$ AGN in the \thirt sample}
The shape of the XLF at high redshift is poorly known because of the difficulties in obtaining a large spectroscopically identified sample of these objects.
We can use the simulated source population to make predictions about the number of high-$z$ AGN in the \thirt sample.
Each of the $f(N_H)$ models predict that around 16\% of the total number of X-ray detections are due to AGN with $3~<~z~<~5$.
Therefore, it can be inferred that the fraction of AGN with $z>3$ in the X-ray population is primarily dependent on the shape of the underlying XLF and its evolution, rather than the $N_H$ distribution within the high-$z$ population.
The model predictions suggest that we should expect around 35 high-$z$ AGN in the \thirt field. 
However, only a single AGN has been identified having $z>3$ by our follow up optical spectroscopy program (which has secure IDs for over 100 sources).
This disparity is maybe due to the over-production of faint sources by the XLF/evolution model; these are more likely to be at high $z$. 
The X-ray detection probability of the $z>3$ AGN is much less dependent on $N_H$ than for the low-$z$ AGN, since most absorption is redshifted below 2 keV.
Therefore, most of the $f(N_H)$ models predict that absorbed AGN make up the majority of the {\em detected} high-$z$ population, the precise fraction being dependent on the particular $f(N_H)$ model.
However, the absorption of optical and UV spectral features does severely affect the probability of identification for these objects.
We have recently obtained further deep optical imaging of the \thirt field in several bands, which will permit us to make photometric redshift estimates for some of the optically faint sources.
The forthcoming deep coverage of the \thirt field in the infrared with \spitzer will further constrain the nature of the high-$z$ population.

\section{Summary}
We have demonstrated how a colour-based analysis of deep \xmm data can be used to constrain models of absorption in the AGN population without requiring complete optical spectroscopic follow up.
By using a detailed simulation technique, we have been able to take account of the complex selection function at work in the sample, and demonstrate how this modulates the input population.
We have shown that a simple $f(N_H)$ model together with an absorbed power-law spectral model (including a reflection component), reproduces the observed $HR1$/$HR2$/$HR3$ colour distribution with probability 6\%. 
All of the other model $N_H$ distributions that we compared were rejected at greater than 99.5\% probablity.
In particular, two more complex $f(N_H)$ models are strongly rejected by the 3D-KS test; the redshift dependent $R=R(z)$ model produces too many hard sources, and the luminosity dependent $R=R(L_X)$ model produces too few.
In general, the addition of a reflection component to the absorbed power-law spectral model improved the match between the colour distributions of the models and the sample.
The reflection component serves to harden the spectral slope at higher energies, and its effect was most evident in the $HR2$ and $HR3$ distributions.
We have shown that there is a large disparity between the shape of the $N(>S_{0.5-2})$ produced by the models and that found in our sample.
We suggest that this is for the most part due to differences between the actual $L-z$ distribution of sources in the \thirt field, and the XLF/evolution model that we have used.
These XLF/evolution differences will have had some effect on the colour distributions produced by the $f(N_H)$ models, and could explain the surfeit of $HR3 = -1$ sources produced by all of the $f(N_H)$ models.
We have seen some evidence that suggests that the spectra of a significant fraction of absorbed sources in the \thirt sample have an additional soft X-ray component.
This feature was not included in our spectral models, and therefore contributed a large part of the disparity between the $HR1$ distributions of models and sample. 
Considering this factor, together with the XLF/evolution differences, we conclude that the 6\% probability for the $\beta =8$ model shows that it provides a rather good fit to the data.
The shape of the $\beta = 8$ distribution can be broadly reproduced using a toy model for the torus in which the density falls away rapidly for viewing angles away from the plane of the torus.
We have shown that AGN having log$N_H > 22$ can be efficiently selected  by choosing sources in the regions $HR1 - \sigma_{HR1} > 0.6$ and $HR2 - \sigma_{HR2} > -0.3$.
We intend to extend the methods described here to further \xmm deep fields in order to increase the sample size, and to reach to fainter X-ray fluxes.

\section*{Acknowledgments}
Based on observations obtained with XMM-Newton, an ESA science mission with instruments and contributions directly funded by ESA Member States and NASA.
TD acknowledges the support of a PPARC Studentship.

\bsp

\label{lastpage}

\end{document}